\documentclass[a4paper, 10pt, conference]{ieeeconf}      

\IEEEoverridecommandlockouts                              
\usepackage{float}
\usepackage{booktabs}
\usepackage{graphicx}
\usepackage{hyperref}
\usepackage{wasysym}
\hypersetup{
    colorlinks = true,
    linkcolor = {blue},
}
\usepackage[normalem]{ulem}
\useunder{\uline}{\ul}{}
\usepackage{caption}

\title{\LARGE \bf
Leveraging Peer Review in Visualization Education: A Proposal for a New Model
}

\author{Alon Friedman$^{1}$ and Paul Rosen$^{2}$
\thanks{*This work was not supported by any organization}
\thanks{$^{1}$A. Friedman is with the School of Information,  University of South Florida,  Tampa, FL 33620, USA
        {\tt\small alon.friedman at usf.edu}}%
\thanks{$^{2}$P. Rosen is with the Department of Computer Science and Engineering, University of South Florida, Tampa, FL 33620, USA
        {\tt\small prosen at usf.edu}}%
}

\begin{document}

\maketitle
\thispagestyle{empty}
\pagestyle{empty}

\begin{abstract}

In visualization education, both science and humanities, the literature is often divided into two parts: the design aspect and the analysis of the visualization. However, we find limited discussion on how to motivate and engage visualization students in the classroom. In the field of Writing Studies, researchers develop tools and frameworks for student peer review of writing. Based on the literature review from the field of Writing Studies, this paper proposes a new framework to implement visualization peer review in the classroom to engage today's students. This framework can be customized for incremental and double-blind review to inspire students and reinforce critical thinking about visualization.

\end{abstract}

\bigskip


\section{INTRODUCTION}

Rushmeier et. al \cite{rushmeier2007revisiting}, defined visual education as a work in progress. Meanwhile, the number of students enrolled in visualization courses, both face-to-face and online, has grown dramatically in recent years---a growth that makes it difficult to provide students the feedback they need to improve the quality of their work. 
%
%

Visualization evaluation has been addressed by a number of researchers \cite{Shamim2014Evaluation, eppler2006comparison, Novak2006Theorigins}. While many techniques and algorithms can be objectively measured, most recognize that there are also significant subjective measurements necessary to completely evaluate the effectiveness of a visualization~\cite{viegas2015design}. Such forms of subjective measurement are not just important for the instructor's assessment of students, but the students also need the ability to critically evaluate a visualization.

For practical purposes, visualization education can be split into two categories. The first is learning the proper construction of visualization---using the right visualization principals in creating your own visualization. This tends to be the primary focus of visualization courses. In an era with increasing proliferation of visualization in science and mass media, the second category is focused on evaluation of the quality and accuracy of visualizations in general. These skills tend to be taught through less formal methods, such as group or class discussion. Less formal methods, while critical, leave students' skills underdeveloped.

To this end, we propose a framework for engaging students in peer review as part of their formal visualization coursework. This framework contains 2 parts. First, we provide a reconfigurable assessment rubric for students to learn how to properly peer review visualizations. Second, we provide a step-by-step approach to educate students in proper evaluation techniques by gradually reducing the constraints on their peer reviews over the course of a semester.

\section{Education and peer review}

Assessment is the heart of formal higher education. Bransford et. al \cite{national2000people} defines assessment as a core component for effective learning. Student assessment has many forms in higher education, and one of these forms, used most commonly in the liberal and creative arts, is peer review. Peer review was established as an essential component for many professional practices, such as the scholarly publication process. The fundamental principle for peer review is that experts in any given domain appraise the professional performance, creativity, or quality of scientific work produced by others in their area of competence. Peer review is often defined as the evaluation of work by one or more people of similar competence to the producers of the work \cite{spier2002history}. The principle component of student peer review is assessment of students by other students, both formative reviews to provide feedback and summative grading.

\section{Moxley Rubric}

\begin{figure}[!t]
	\centering
	\includegraphics[trim = 0 6pt 0 0, clip, width=1.0\linewidth]{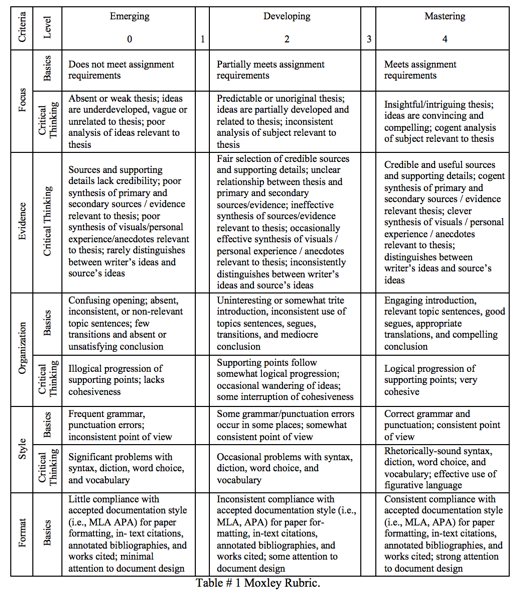}
	\caption{Reproduction of the Moxley Rubric from~\cite{moxley2013big}.}
	\label{fig:moxley}
\end{figure}

Moxley \cite{moxley2013big} reports that the use of peer review technologies is changing the ecology of assessment and student motivation. ``Rather than happening after the fact, online assessment systems are becoming part of the teaching and learning process, part of the digital experience that students are so motivated to be part of.'' Moxley utilizes this rubric to develop online peer review, where the instructor can choose between two versions of the common rubric: (1) the numeric rubric, which requires students to score rubric criteria (Focus, Evidence, Organization, Style, and Format) based on a five-point scale; and (2) the discussion rubric, which requires students to write textual comments regarding these criteria rather than scores. While the textual feedback provides more detail, research has shown that instructors have favored the numeric version of the rubric over the discussion version \cite{moxley2013big}. A reproduction of the rubric is seen in Figure~\ref{fig:moxley}.

\section{Background in Visualization Education}

In 2006, the ACM Siggraph Education Committee started the discussion on what visualization education should be. Domik~\cite{domik2000we} recaptured this discussion by posting the question: ``Do we need formal education in visualization?'' She answered this question by stating that we do at least need informal visualization education. In the 2000s, humanities departments started to offer visualization courses to their students, and as a result, Pat Hanrahan proposed implementation of core topics to fit the humanities, including:  data and image models; perception and cognition; interaction; space; color; shape and lines; texture; interior structure with volumetric techniques; and narrative with animation~\cite{ma2005teaching}. 

For this study, we embraced Hanrahan's core topics in our classes, due in part to the different backgrounds our students bring to our classes. Understanding the breadth of these topic areas is not necessarily sufficient to be considered skilled in visualization. Gilbert \cite{gilbert2008visualization} described 5 levels of competency that give insight to the depth of topical understanding in visualization. His method outlines step-by-step instruction on visualization design that includes: step 1, representation as depiction; step 2, converting early symbolic skills; step 3, the syntactic use of formal representation; step 4, semantic use of formal representations; and step 5, reflective, rhetorical use of presentation. However, we did not find any discussion or references on how to embed these steps into the analysis of visualization or peer review of visualization.

\section{Approach}

In 2013 and 2014, the University of South Florida hired two faculty into the College of Arts and Sciences and College of Engineering, respectively, to build visualization into their curriculum. This effort was taking place in the context of a larger university-led effort towards improving visualization skills and access to visualization equipment. Both faculty focus on teaching visualization as any technique for creating images, diagrams, or animations to communicate a message using different platforms and programming languages. 

In the  School of Information under the College of Arts and Sciences, a course called ``Visualization of Big Data'' was offered. The class is part of new area of concentration in Data Science and was offered to students via an online setting only. In the College of Engineering the Department of Computer Science and Engineering offered the class called ``Scientific Visualization.'' The class was offered in traditional classroom format only as an elective. The key differences between these courses were format (i.e., online vs.\ classroom) and emphasis (i.e., design for the Information Science course and algorithmic for the Computer Science course). In both classes the instructors focused not just on design and algorithmic technique, but they also focused on the analysis of visualizations using models such as the ``what, why, and how'' model of Liu and Stasko \cite{liu2010theories} as a framework for discussion.

During these 2 courses, we employed several peer review techniques. We collected the students' peer review comments with the hope to find patterns. Our premises for the investigation was based on the three principals of Monk et al.~\cite{monk2002funology}:

\begin{itemize}
   \item Careless mapping from data to pictures may lead to erroneous interpretation.
   \item A substantial amount of knowledge is necessary to generate images depicting complex information in a way that prevents erroneous interpretation.
   \item Decision making is increasingly based on visual representations.
\end{itemize}

\section{Results}

We did not find a single student who directly refereed to the Monk's principles in their peer review, as they were covered in class and in their assignments. Many of the students selected one aspect of visualization without considering the other aspects that are involved in creating and evaluating visualization work. For example, many students simply ignored considerations such as the variables embedded, visual patterns, or time. To illustrate our findings, the following is a common example from a student peer review. The student in this case was asked to review their peer's histogram created using R, based on Monk's principles. The histogram had to be displayed in multiple colors as part of the assignment.

\vspace{10pt}
\noindent\hspace{0.075\linewidth}\begin{minipage}{0.85\linewidth}
\emph{This is the first time in my life I had to provide comments and criticism on my peers’ work using open-source R. I certainly had my struggles with R to create visualization, but in this assignment, I have problems understanding to the complexity of visualization and visualization creation using R.}
\end{minipage}
\vspace{10pt}

Furthermore, many of the comments we examined did not have any direct connection to the content of the visualization itself. Thus, we propose a new framework to better teach the evaluation of visualizations.

\section{Proposed Framework}

In the field of Writing Studies, researchers have reported on the practice of using generic rubrics to make global claims about writing development. Beaufort \cite{beaufort2008college}, Nowacek \cite{Nowacek2011Agents}, and Wardle and Roozen \cite{wardle2012addressing} have explored how the ability to write well is a skill that can be mastered in one context and then simply carried over to another context. Moxley \cite{moxley2013big} opposed this approach stating that arguments that make grand claims about student ability based on a handful of rubric scores need to be seriously challenged. Students' scores on one rubric are not necessarily predictive of how they will do when facing alternative genres. To address the limited availability of visualization peer review tools and techniques in the classroom, we propose the following framework adhering to Moxley's recommendation.

\subsection{Peer Review Rubric for visualization education}

The basic structure of our rubric is to divide topics into 5 major assessment categories, with each category having sub-assessments. The sub-assessments contain 3 scoring categories affixed to a 5-point scale (though no scale is shown). The category on the left is reserved for poor performance on the assessment, middle for average performance, and right for excellent performance. Each sub-assessment contains a comments box for providing details about the scoring.

For our rubric, the 5 major assessment categories are: algorithm design, interaction design, visual design, design consideration, and visualization narrative. The categories cover the following subjects: The algorithm design category is concerned with algorithm selection and implementation. Interaction design is concerned with how the user interacts with the visualization. Visual design is concerned with the technical aspects of how data are placed in the visualization. Examples include, visual encoding channels, their expressiveness, and their effectiveness. Design consideration focuses on composition and aesthetics aspects of the visualization, such as embellishments. The final category visualization narrative is critical in projects where the story surrounding the visualization is as important as the visualization itself. This category provides a basic framework for assessing whether the visualization supports the story and vice versa.

\subsection{Customize Rubric}

We also consider customization of the rubric, so it can be built based upon the content of an assignment or customized to fit the course content. Instructors should take our full template, extract the critical components, and add missing components with respect to their assignment and course content. For example, on a computer science assignment, the rubric provides the ability to remove or reduce the number of components dealing with aesthetics and narrative, while emphasizing those for algorithm. On the other hand, for projects that do not implement an algorithm or those for which the algorithm is not critical may skip this category. 

Another aspect in which this rubric can also be customized is to reduce the constraints of the rubric itself. Assuming the critical thinking skills of students, particularly in the topic of visualization, are weak, the rubric itself can be a tool to help improve those skills. At the beginning of the course, the rubric can be provided that includes all scoring categories to build an analytic framework in the students' minds. As their skills improve, the 5-point scoring scale can be removed, leaving only the comments section behind. Finally, as students begin to perfect their skills, the sub-assessments can be removed altogether, leaving only the major categories of evaluation. In this way, students will go from highly constrained and guided evaluation to free-form evaluation over the course of a semester.

Finally, we foresee the peer review process appearing much like conference or journal reviews. First, double-blind review helps to reduce the risk of collusion or malicious review. Second, multiple reviews per visualization help to ensure that consensus is built around the evaluation. In addition, each student having to produce multiple reviews helps to reinforce the critical thinking skills. Some of these functionalities are available in learning management systems, such as Canvas.

\section{Risks in Peer Review}

As we implement this framework in our own classrooms, we are keeping an eye out for a couple of key issues. For example, peer review is a compliment to existing educational approaches, but the instructor, as the expert, still needs to provide feedback. Otherwise, some important subtleties may be missed. Second, the risk for collusion or malice are great, even with double-blind reviews. Students talk, and when they talk, they will undoubtedly discover they are evaluating each other's visualizations. Therefore, instructors should take care to randomize peer reviewers, and grades should only be loosely based upon the results of the peer review. It is also important to remember that there is both art and science to design with no single optimal design. Therefore, a peer review may not necessarily be the best mechanism for providing grades. However, peer review naturally supports iterative design improvements~\cite{viegas2015design}.

\section{Conclusion}

We have presented a new framework for evaluating student visualizations based upon peer review rubric. This rubric has a couple of important aspects. First, it provided students a framework for critical thinking and evaluation of visualizations. Second, it provides a mechanism to help with issues of larger class sizes, while still providing students feedback about their work.

Finally, we do not believe that the rubric as presented will be the final static version. We anticipate it will be a growing and evolving document as community members provide their input and the focus of the visualization community changes. As a result,  Appendices A-E represents our proposal for student peer review rubrics based on specific subjects from algorithmic design, interaction design, visual design, design consideration, and  visualization narrative.  For the full proposed visualization rubric, see Appendix A-E. A \LaTeX version of the rubric can be cloned/forked at \url{https://github.com/USFDataVisualization/VisPeerReview}.

\section*{APPENDIX}

\begin{table}[H]
\centering
\caption*{\textbf{Appendix A}: Algorithmic Design }
\label{t1}
\resizebox{\columnwidth}{!}{%
\begin{tabular}{|l|ccccc|}
\hline
	\begin{tabular}[c]{@{}l@{}}\uline{Selection of Algorithm} -   Of the \\ available algorithm options was the best \\ selected?  What aspects of this algorithm \\  make it better or worse than the \\  other possible choices?\end{tabular} 
	& \multicolumn{1}{c|}{\Square \ \begin{tabular}[c]{@{}c@{}}Below\\ Average\end{tabular}} 
    & \multicolumn{1}{c|}{\Square} 
    & \multicolumn{1}{c|}{\Square \ Average} 
    & \multicolumn{1}{c|}{\Square} 
    & \Square \ \begin{tabular}[c]{@{}c@{}}Above\\ Average\end{tabular} \\ 
    \cline{2-6} 
 	& Comments: &  &  &  &  \\ \hline
 
\begin{tabular}[c]{@{}l@{}} \uline{Correct Implementation} - Does the \\ algorithm appear to produce the correct \\ result, given your knowledge of the data?\end{tabular} 
	& \multicolumn{1}{c|}{\Square \ No} 
    & \multicolumn{1}{c|}{\Square} 
    & \multicolumn{1}{c|}{\Square \ \begin{tabular}[c]{@{}c@{}}Minor\\ Errors\end{tabular}} 
    & \multicolumn{1}{c|}{\Square} 
    & \Square \ \begin{tabular}[c]{@{}c@{}}Appears\\ Correct\end{tabular} \\ 
    \cline{2-6} 
 	& Comments: &  &  &  &  \\ \hline
 
	\begin{tabular}[c]{@{}l@{}} \uline{Efficient Implementation} - Is the \\ performance (speed) of the algorithm what \\ you expected? Is is slower? Is it faster?\end{tabular} 
    & \multicolumn{1}{c|}{\Square \ \begin{tabular}[c]{@{}l@{}}Much \\ Slower\end{tabular}} 
    & \multicolumn{1}{c|}{\Square} 
    & \multicolumn{1}{c|}{\Square \ \begin{tabular}[c]{@{}l@{}}As \\ Expected\end{tabular}} 
    & \multicolumn{1}{c|}{\Square} 
    & \Square \ \begin{tabular}[c]{@{}l@{}}Much \\ Faster\end{tabular} \\ 
    \cline{2-6} 
 	& Comments: &  &  &  &  \\ \hline
    
	\begin{tabular}[c]{@{}l@{}} \uline{Featureful Implementation} - Does the \\ implementation contain the basic \\  required features or are additional features \\ included?\end{tabular} 
    & \multicolumn{1}{c|}{\Square \ \begin{tabular}[c]{@{}l@{}}Major \\ Features \\ Missing\end{tabular}} 
    & \multicolumn{1}{c|}{\Square} 
    & \multicolumn{1}{c|}{\Square \ \begin{tabular}[c]{@{}l@{}}As \\ Expected\end{tabular}} 
    & \multicolumn{1}{c|}{\Square} 
    & \Square \ \begin{tabular}[c]{@{}l@{}}Major \\ Features \\ Added\end{tabular} \\ 
    \cline{2-6} 
 	& Comments: &  &  &  &  \\ \hline
    
	\begin{tabular}[c]{@{}l@{}} \uline{Datasets Used} - Do the datasets provided \\ give enough information to evaluate the \\ correctness, efficiency, and featurefulness \\ of the implementation?\end{tabular} 
    & \multicolumn{1}{c|}{\Square \ \begin{tabular}[c]{@{}l@{}}Not \\ Useful\end{tabular}} 
    & \multicolumn{1}{c|}{\Square} 
    & \multicolumn{1}{c|}{\Square \ \begin{tabular}[c]{@{}l@{}}As \\ Expected\end{tabular}} 
    & \multicolumn{1}{c|}{\Square} 
    & \Square \ \begin{tabular}[c]{@{}l@{}}Better \\ Than \\ Expected\end{tabular} \\ 
    \cline{2-6} 
 	& Comments: &  &  &  &  \\ \hline
\end{tabular}%
}
\end{table}

\begin{table}[H]
\centering
\caption*{\textbf{Appendix B}: Interaction Design}
\label{t2}
\resizebox{\columnwidth}{!}{%
\begin{tabular}{|llllll|}
\hline
\multicolumn{1}{|l|}{\begin{tabular}[c]{@{}l@{}}\uline{Interaction Selection} - What \\ interaction mechanisms are \\ being used?\end{tabular}} 
	& \multicolumn{5}{|l|}{
        \begin{tabular}[c]{lll}  
  			\Square \ Linked Views  & \Square \ Filtering  	  & \Square \ Geometric Zoom	 \\ 
			\Square \ Selection		& \Square \ Aggregation	  & \Square \ Pan/Translate	 \\ 
  			\Square \ Highlighting 	& \Square \ Semantic Zoom & \Square \ Rotate 			 \\
        \end{tabular}
  	}
	\\ \cline{2-6}
	\multicolumn{1}{|l|}{} & Comments: &  &  &  &  \\ \hline
    
	\multicolumn{1}{|l|}{\begin{tabular}[c]{@{}l@{}}\uline{Interaction Effectiveness} - Are \\ the interactions provided highly \\ effective? In other words, does \\ the visualization react in a \\ manner that makes it easy to \\ use or capable of providing rich \\ content?\end{tabular}} 
    & \multicolumn{1}{l|}{\Square \ \begin{tabular}[c]{@{}l@{}}Missing \\ Key\\  Interactions\end{tabular}} 
    & \multicolumn{1}{l|}{\Square} 
    & \multicolumn{1}{l|}{\Square \ Expected} 
    & \multicolumn{1}{l|}{\Square} 
    & \Square \ \begin{tabular}[c]{@{}l@{}}Better \\ Than \\ Expected\end{tabular} \\ \cline{2-6} 
\multicolumn{1}{|l|}{} & Comments: &  &  &  &  \\ \hline
\end{tabular}%
}
\end{table}

\begin{table}[H]
\centering
\caption*{\textbf{Appendix C}: Visual Design}
\label{t3}
\resizebox{\columnwidth}{!}{%
\begin{tabular}{|llllll|}
\hline
  \multicolumn{1}{|l|}{\begin{tabular}[c]{@{}l@{}}\uline{Visual Encodings} - What visual channels \\ were used to encode data?\end{tabular}} 
	& \multicolumn{5}{|l|}{
        \begin{tabular}[c]{llll}
  			\Square \ Position  & \Square \ Curvature	& \Square \ Area	& \Square \ Color Hue  \\ 
			\Square \ Depth		& \Square \ Shape		& \Square \ Volume	& \Square \ Texture  \\ 
  			\Square \ Angle 	& \Square \ Length		& \Square \ \begin{tabular}[c]{@{}l@{}}Luminance/\\ Saturation\end{tabular} & \Square \ \begin{tabular}[c]{@{}l@{}}Motion /\\ Animation\end{tabular}  
  	\end{tabular}
    }
  \\ \cline{2-6} 
  \multicolumn{1}{|l|}{} & Comments: &  &  &  &  \\ \hline
  
  \multicolumn{1}{|l|}{\begin{tabular}[c]{@{}l@{}}\uline{Intended/Unintended Encodings} - Do all \\ of the visual encoding appear to be \\ intended, or were some accidentally \\ created?\end{tabular}} 
  & \multicolumn{1}{l|}{\Square \ \begin{tabular}[c]{@{}l@{}}Many \\ Unintended\end{tabular}} 
  & \multicolumn{1}{l|}{\Square} 
  & \multicolumn{1}{l|}{\Square \ \begin{tabular}[c]{@{}l@{}}Few \\ Unintended\end{tabular}} 
  & \multicolumn{1}{l|}{\Square} 
  & \Square \ \begin{tabular}[c]{@{}l@{}}All \\ Intended\end{tabular} \\ \cline{2-6} 
  \multicolumn{1}{|l|}{} & Comments: &  &  &  &  \\ \hline
  
  \multicolumn{1}{|l|}{\begin{tabular}[c]{@{}l@{}}\uline{Expressiveness of Encodings} - Are the \\ visual encodings attached to the correct \\ type of data for that encoding (i.e. are \\ quantitative data attached to quantitative \\ encodings and categorical data to \\ categorical encodings)?\end{tabular}} 
  & \multicolumn{1}{l|}{\Square \ \begin{tabular}[c]{@{}l@{}}Many \\ Errors\end{tabular}} 
  & \multicolumn{1}{l|}{\Square} 
  & \multicolumn{1}{l|}{\Square \ \begin{tabular}[c]{@{}l@{}}Few \\ Errors\end{tabular}} 
  & \multicolumn{1}{l|}{\Square} 
  & \Square \ \begin{tabular}[c]{@{}l@{}}Correctly \\ Assigned\end{tabular} \\ \cline{2-6} 
  \multicolumn{1}{|l|}{} & Comments: &  &  &  &  \\ \hline
  
  \multicolumn{1}{|l|}{\begin{tabular}[c]{@{}l@{}}\uline{Effectiveness of Encodings} - Have the \\ maximally effective visual encodings \\ been selected in all cases?\end{tabular}} 
  & \multicolumn{1}{l|}{\Square \ \begin{tabular}[c]{@{}l@{}}Many \\ Ineffective\end{tabular}} 
  & \multicolumn{1}{l|}{\Square} 
  & \multicolumn{1}{l|}{\Square \ \begin{tabular}[c]{@{}l@{}}Few \\ Ineffective\end{tabular}} 
  & \multicolumn{1}{l|}{\Square} 
  & \Square \ \begin{tabular}[c]{@{}l@{}}Most \\ Effective\end{tabular} \\ \cline{2-6} 
  \multicolumn{1}{|l|}{} & Comments: &  &  &  &  \\ \hline
  
  \multicolumn{1}{|l|}{\begin{tabular}[c]{@{}l@{}}\uline{Integral vs. Separable (Conjunction)} - \\ When visual encodings are mixed, do \\ the combined encodings make the \\ visualization more effective or do they  \\ make interpretation more difficult?\end{tabular}} 
  & \multicolumn{1}{l|}{\Square \ \begin{tabular}[c]{@{}l@{}}Mostly \\ Ineffective\end{tabular}} 
  & \multicolumn{1}{l|}{\Square} 
  & \multicolumn{1}{l|}{\Square \ \begin{tabular}[c]{@{}l@{}}None \\ Used\end{tabular}} 
  & \multicolumn{1}{l|}{\Square} 
  & \Square \ \begin{tabular}[c]{@{}l@{}}Highly \\ Effective\end{tabular} \\ \cline{2-6} 
  \multicolumn{1}{|l|}{} & Comments: &  &  &  &  \\ \hline
  
  \multicolumn{1}{|l|}{\begin{tabular}[c]{@{}l@{}}\uline{Effective Use of Color} - Is color used \\ in a same fashion? Do the colors chosen \\ and the application of those colors make \\ the visualization effective?\end{tabular}} 
  & \multicolumn{1}{l|}{\Square \ \begin{tabular}[c]{@{}l@{}}Mostly \\ Ineffective\end{tabular}} 
  & \multicolumn{1}{l|}{\Square} 
  & \multicolumn{1}{l|}{\Square \ \begin{tabular}[c]{@{}l@{}}None \\ Used\end{tabular}} 
  & \multicolumn{1}{l|}{\Square} 
  & \Square \ \begin{tabular}[c]{@{}l@{}}Highly \\ Effective\end{tabular} \\ \cline{2-6} 
  \multicolumn{1}{|l|}{} & Comments: &  &  &  &  \\ \hline
  
  \multicolumn{1}{|l|}{\begin{tabular}[c]{@{}l@{}}\uline{Color Contrast and Harmony} - Do the \\ selected colors properly contrast and \\ harmonize?\end{tabular}} 
  & \multicolumn{1}{l|}{\Square \ No} 
  & \multicolumn{1}{l|}{\Square} 
  & \multicolumn{1}{l|}{\Square \ \begin{tabular}[c]{@{}l@{}}No \\ Color \\ Used\end{tabular}} 
  & \multicolumn{1}{l|}{\Square} 
  & \Square \ Yes \\ \cline{2-6} 
  \multicolumn{1}{|l|}{} & Comments: &  &  &  &  \\ \hline
  
  \multicolumn{1}{|l|}{\begin{tabular}[c]{@{}l@{}}\uline{Colorblind Safety} - Is the visualization \\ colorblind safe? Were redundant visual \\ encodings used?\end{tabular}} 
  & \multicolumn{1}{l|}{\Square \ No} 
  & \multicolumn{1}{l|}{\Square} 
  & \multicolumn{1}{l|}{\Square \ \begin{tabular}[c]{@{}l@{}}Colorblind \\ Safe \\ Color\end{tabular}} 
  & \multicolumn{1}{l|}{\Square} 
  & \Square \ Redundant \\ \cline{2-6} 
  \multicolumn{1}{|l|}{} & Comments: &  &  &  &  \\ \hline
\end{tabular}%
}
\end{table}

\begin{table}[H]
\centering
\caption*{\textbf{Appendix D}: Design Consideration}
\label{t4}
\resizebox{\columnwidth}{!}{%
\begin{tabular}{|llllll|}
\hline
  \multicolumn{1}{|l|}{\begin{tabular}[c]{@{}l@{}}Clear, Detailed, and Thorough \\ Labeling - Is appropriate and \\ complete labeling used throughout \\ or do missing labels require \\ assumptions about the data?\end{tabular}} 
  & \multicolumn{1}{l|}{\Square \ \begin{tabular}[c]{@{}l@{}}No \\ Labels\end{tabular}} 
  & \multicolumn{1}{l|}{\Square} 
  & \multicolumn{1}{l|}{\Square \ \begin{tabular}[c]{@{}l@{}}Some \\ Missing \\ Labels\end{tabular}} 
  & \multicolumn{1}{l|}{\Square} 
  & \Square \ \begin{tabular}[c]{@{}l@{}}Completely \\ Labeled\end{tabular} \\ \cline{2-6} 
  \multicolumn{1}{|l|}{} & Comments: &  &  &  &  \\ \hline
  
  \multicolumn{1}{|l|}{\begin{tabular}[c]{@{}l@{}}Missing Scales - Are scales \\ provided for the data?\end{tabular}} & \multicolumn{1}{l|}{\Square \ \begin{tabular}[c]{@{}l@{}}No \\ Scales\end{tabular}} 
  & \multicolumn{1}{l|}{\Square} 
  & \multicolumn{1}{l|}{\Square \ \begin{tabular}[c]{@{}l@{}}Some \\ Missing \\ Scales\end{tabular}} 
  & \multicolumn{1}{l|}{\Square} 
  & \Square \ \begin{tabular}[c]{@{}l@{}}All \\ Scales \\ Present\end{tabular} \\ \cline{2-6} 
  \multicolumn{1}{|l|}{} & Comments: &  &  &  &  \\ \hline
  
  \multicolumn{1}{|l|}{\begin{tabular}[c]{@{}l@{}}Missing Legend - Is a legend \\ provided for the data? Does the \\ legend provide useful information?\end{tabular}} 
  & \multicolumn{1}{l|}{\Square \ \begin{tabular}[c]{@{}l@{}}No \\ Legend\end{tabular}} 
  & \multicolumn{1}{l|}{\Square} 
  & \multicolumn{1}{l|}{\Square \ \begin{tabular}[c]{@{}l@{}}Incomplete \\ Legend\end{tabular}} 
  & \multicolumn{1}{l|}{\Square} 
  & \Square \ \begin{tabular}[c]{@{}l@{}}Complete \\ Legend\end{tabular} \\ \cline{2-6} 
  \multicolumn{1}{|l|}{} & Comments: &  &  &  &  \\ \hline
  
  \multicolumn{1}{|l|}{\begin{tabular}[c]{@{}l@{}}Scale Distortion - Is any scale \\ distortion or deception used in \\ the visualization?\end{tabular}} 
  & \multicolumn{1}{l|}{\Square \ \begin{tabular}[c]{@{}l@{}}Severe \\ Distortion\end{tabular}} 
  & \multicolumn{1}{l|}{\Square} 
  & \multicolumn{1}{l|}{\Square \ \begin{tabular}[c]{@{}l@{}}Minor \\ Distortion\end{tabular}} 
  & \multicolumn{1}{l|}{\Square} 
  & \Square \ \begin{tabular}[c]{@{}l@{}}No Distortion\end{tabular} \\ \cline{2-6} 
  \multicolumn{1}{|l|}{} & Comments: &  &  &  &  \\ \hline
  
  \multicolumn{1}{|l|}{\begin{tabular}[c]{@{}l@{}}Lie Factor - Is there any lie factor? \\ How extreme is the lie factor?\end{tabular}} 
  & \multicolumn{1}{l|}{\Square \ \begin{tabular}[c]{@{}l@{}}Major \\ Lie\end{tabular}} 
  & \multicolumn{1}{l|}{\Square} 
  & \multicolumn{1}{l|}{\Square \ \begin{tabular}[c]{@{}l@{}}Minor \\ Lie\end{tabular}} 
  & \multicolumn{1}{l|}{\Square} 
  & \Square \ \begin{tabular}[c]{@{}l@{}}No \\ Lie\end{tabular} \\ \cline{2-6} 
  \multicolumn{1}{|l|}{} & Comments: &  &  &  &  \\ \hline
  
  \multicolumn{1}{|l|}{\begin{tabular}[c]{@{}l@{}}Data/Ink Ratio - Is the data to ink \\ ratio reasonable? Could it be more \\ efficient?\end{tabular}} 
  & \multicolumn{1}{l|}{\begin{tabular}[c]{@{}l@{}}\Square \ \begin{tabular}[c]{@{}l@{}}Way Too \\ Little /\end{tabular} \\ \Square \ Much Ink\end{tabular}} 
  & \multicolumn{1}{l|}{\Square} 
  & \multicolumn{1}{l|}{\begin{tabular}[c]{@{}l@{}}\Square \ \begin{tabular}[c]{@{}l@{}}Slightly Too \\ Little /\end{tabular} \\ \Square \ Much Ink\end{tabular}} 
  & \multicolumn{1}{l|}{\Square} 
  & \Square \ \begin{tabular}[c]{@{}l@{}}Perfect \\ Amount \\ of Ink\end{tabular} \\ \cline{2-6} 
  \multicolumn{1}{|l|}{} & Comments: &  &  &  &  \\ \hline
  
  \multicolumn{1}{|l|}{\begin{tabular}[c]{@{}l@{}}Chart Junk, Embellishments, \\ Aesthetics - Are appropriate \\ embellishments used? Are the \\ embellishments distracting? Do \\ the embellishments add to \\ the visualization?\end{tabular}} 
  & \multicolumn{1}{l|}{\begin{tabular}[c]{@{}l@{}}\Square \ \begin{tabular}[c]{@{}l@{}} Way Too Few /\end{tabular} \\\Square \ \begin{tabular}[c]{@{}l@{}} Many \\ Embellishments\end{tabular}\end{tabular}} 
  & \multicolumn{1}{l|}{\Square} 
  & \multicolumn{1}{l|}{\begin{tabular}[c]{@{}l@{}}\Square \ \begin{tabular}[c]{@{}l@{}} A Bit Too \\ Few /\end{tabular} \\\Square \ \begin{tabular}[c]{@{}l@{}} Many \\ Embellishments\end{tabular}\end{tabular}} 
  & \multicolumn{1}{l|}{\Square} 
  & \Square \ \begin{tabular}[c]{@{}l@{}}Perfect \\ Number of \\ Embellishments\end{tabular} \\ \cline{2-6} 
  \multicolumn{1}{|l|}{} & Comments: &  &  &  &  \\ \hline
  
  \multicolumn{1}{|l|}{\begin{tabular}[c]{@{}l@{}}Data Density - Has too much data \\ been included in the visualization \\ making interpretation difficult?\end{tabular}} 
  & \multicolumn{1}{l|}{\Square \ \begin{tabular}[c]{@{}l@{}}Too \\Sparse\end{tabular}} 
  & \multicolumn{1}{l|}{\Square} 
  & \multicolumn{1}{l|}{\Square \ \begin{tabular}[c]{@{}l@{}}Expected\end{tabular}} 
  & \multicolumn{1}{l|}{\Square} 
  & \Square \ \begin{tabular}[c]{@{}l@{}}Too \\Dense\end{tabular} \\ \cline{2-6} 
  \multicolumn{1}{|l|}{} & Comments: &  &  &  &  \\ \hline
  
  \multicolumn{1}{|l|}{\begin{tabular}[c]{@{}l@{}}Task Selection - Does the \\ visualization enable appropriate \\ visual analysis tasks for the data \\ type and/or dataset?\end{tabular}} 
  & \multicolumn{1}{l|}{\Square \ \begin{tabular}[c]{@{}l@{}}Unclear \\ Task \\ Selection\end{tabular}} 
  & \multicolumn{1}{l|}{\Square}
  & \multicolumn{1}{l|}{\Square \ \begin{tabular}[c]{@{}l@{}}Some \\ Tasks \\ Missing\end{tabular}} 
  & \multicolumn{1}{l|}{\Square} 
  & \Square \ \begin{tabular}[c]{@{}l@{}}Wide Range \\ of Tasks \\ Support\end{tabular} \\ \cline{2-6} 
  \multicolumn{1}{|l|}{} & Comments: &  &  &  &  \\ \hline
  
  \multicolumn{1}{|l|}{\begin{tabular}[c]{@{}l@{}}Gestalt Principals - Have Gestalt \\ principals been used to improve \\ analysis?\end{tabular}} 
  & \multicolumn{1}{l|}{\Square \ \begin{tabular}[c]{@{}l@{}}No \\ Gestalt \\ Principals\end{tabular}} 
  & \multicolumn{1}{l|}{\Square} 
  & \multicolumn{1}{l|}{\Square \ \begin{tabular}[c]{@{}l@{}}Some \\ Gestalt \\ Principals\end{tabular}} 
  & \multicolumn{1}{l|}{\Square} 
  & \Square \ \begin{tabular}[c]{@{}l@{}}Many \\ Gestalt \\ Principals\end{tabular} \\ \cline{2-6} 
  \multicolumn{1}{|l|}{} & Comments: &  &  &  &  \\ \hline
\end{tabular}%
}
\end{table}

\begin{table}[H]
\centering
\caption*{\textbf{Appendix E}: Visualization Narrative}
\label{t5}
\resizebox{\columnwidth}{!}{%
\begin{tabular}{|llllll|}
\hline
  \multicolumn{1}{|l|}{\begin{tabular}[c]{@{}l@{}}Description of Visualization - Is \\ a description of  the visualization \\ accurate and informative?\end{tabular}} 
  & \multicolumn{1}{l|}{\Square \ \begin{tabular}[c]{@{}l@{}}No \\ Description\end{tabular}} 
  & \multicolumn{1}{l|}{\Square} 
  & \multicolumn{1}{l|}{\begin{tabular}[c]{@{}l@{}}\Square \ Incomplete / \\ \Square \ Self-Explanatory\end{tabular}} 
  & \multicolumn{1}{l|}{\Square} 
  & \Square \ \begin{tabular}[c]{@{}l@{}}Complete \\ Description\end{tabular} \\ \cline{2-6} 
  \multicolumn{1}{|l|}{} & Comments: &  &  &  &  \\ \hline
  
  \multicolumn{1}{|l|}{\begin{tabular}[c]{@{}l@{}}Support of Narrative - Does the \\ visualization support the message \\ of the narrative?\end{tabular}} 
  & \multicolumn{1}{l|}{\Square \ \begin{tabular}[c]{@{}l@{}}No \\ Description\end{tabular}} 
  & \multicolumn{1}{l|}{\Square} 
  & \multicolumn{1}{l|}{\begin{tabular}[c]{@{}l@{}}\Square \ Incomplete / \\ \Square \ Self-Explanatory\end{tabular}} 
  & \multicolumn{1}{l|}{\Square} 
  & \Square \ \begin{tabular}[c]{@{}l@{}}Completely \\ Supportive\end{tabular} \\ \cline{2-6} 
  \multicolumn{1}{|l|}{} & Comments: &  &  &  &  \\ \hline
  
  \multicolumn{1}{|l|}{\begin{tabular}[c]{@{}l@{}}Dataset Used - Do the dataset(s) \\ provide enough information and \\ detail to support the narrative?\end{tabular}} 
  & \multicolumn{1}{l|}{\Square \ \begin{tabular}[c]{@{}l@{}}Not At \\ All\end{tabular}} 
  & \multicolumn{1}{l|}{\Square} 
  & \multicolumn{1}{l|}{\Square \ Partially} 
  & \multicolumn{1}{l|}{\Square} 
  & \Square \ Completely \\ \cline{2-6} 
  \multicolumn{1}{|l|}{} & Comments: &  &  &  &  \\ \hline
\end{tabular}%
}
\end{table}

\bibliographystyle{IEEEtran}
\bibliography{IEEEabrv,main}

\end{document}